\documentclass{ws-procs975x65} 
\usepackage{graphics,color}
\usepackage[sort&compress]{natbib}
\def\TrTrOne{ $SU(3)_C$ $\otimes$ $SU(3)_L$ $\otimes$ $U(1)_X$ }
\usepackage{graphicx}

\newcommand{\bad}{\begin{array}{ccc}}
\DeclareMathAlphabet{\mathsc}{OT1}{cmr}{m}{sc}

\usepackage{color}

\def\be{\begin{equation}}        
\def\ee{\end{equation}} 
\def\bear{\be\begin{array}}       
\def\eear{\end{array}\ee} 
\def\bea{\begin{eqnarray}} 
\def\eea{\end{eqnarray}}

\def\beqa{\begin{eqnarray}}
\def\eeqa{\end{eqnarray}}

\def\beq{\begin{equation}}
\def\eeq{\end{equation}}
\def\ba{\begin{array}}
\def\ea{\end{array}}

\def\bold#1{\setbox0=\hbox{$#1$} 
     \kern-.025em\copy0\kern-\wd0 
     \kern.05em\copy0\kern-\wd0 
     \kern-.025em\raise.0433em\box0 }

\def\lfv{lepton flavour violation }
\def\lnv{lepton number violation }

\newcommand {\ignore}[1]{}

\def\U1{$\mathrm{U(1)}$ }
\def\e6{$\mathrm{E(6)}$ }
\def\10{$\mathrm{SO(10)}$ }
\def\21{$\mathrm{SU(2)_L \otimes U(1)_Y}$ }

\def\31{$\mathrm{SU(3)_c \otimes U(1)_Q}$ }
\def\SM{$\mathrm{SU(3)_c \otimes SU(2)_L \otimes U(1)_Y}$ }

\newcommand{\sm}{{standard model }}

\def\3211{$\mathrm{SU(3) \otimes SU(2)_L \otimes U(1)_R \otimes U(1)_{B-L}}$ }
\def\321{$\mathrm{SU(3) \otimes SU(2) \otimes U(1)}$ }
\def\422{$\mathrm{SU(4) \otimes SU(2) \otimes SU(2)_R}$ }

\def \znbb {$\rm 0\nu\beta\beta$ }

\def\lsim{\mathrel{\rlap{\lower4pt\hbox{\hskip1pt$\sim$}}
    \raise1pt\hbox{$<$}}}         
\def\gsim{\mathrel{\rlap{\lower4pt\hbox{\hskip1pt$\sim$}}
    \raise1pt\hbox{$>$}}}         

\makeatletter
\renewcommand{\fnum@table}{\textbf{\tablename~\thetable}}
\renewcommand{\fnum@figure}{\textbf{\figurename~\thefigure}}
\makeatother

\def\mxth{\mathsurround=0pt }
\def\xversim#1#2{\lower2.pt\vbox{\baselineskip0pt \lineskip-.5pt
  \ialign{$\mxth#1\hfil##\hfil$\crcr#2\crcr\sim\crcr}}}

\def\be{\begin{equation}}
\def\ee{\end{equation}}
\def\bea{\begin{eqnarray}}
\def\eea{\end{eqnarray}}

\begin{document}
\thispagestyle{empty}
\noindent\
\\
\\
\\
\begin{center}
\large \bf 
\end{center}
\hfill
 \vspace*{1cm}
\noindent
\begin{center}
\title{Status and implications of neutrino masses:\\
a brief panorama}
\author{\bf Jos\'e W.F. Valle$^*$}
\address{Instituto de F\'{\i}sica Corpuscular (CSIC-UV)\\
Parc Cient\'ific de la Universitat de Val\`{e}ncia\\
C/ Catedr\'atico Jos\'e Beltr\'an, 2\\
E-46980 Paterna, Valncia, Spain,\\
$^*$ https://www.astroparticles.es/}
\vspace*{0.5cm}
\end{center}

\begin{abstract}

  With the historic discovery of the Higgs boson our picture of
  particle physics would have been complete were it not for the neutrino
  sector and cosmology.
  I briefly discuss the role of neutrino masses and mixing upon gauge
  coupling unification, electroweak breaking and the flavor sector.
  Time is ripe for new discoveries such as leptonic CP violation,
  charged lepton flavor violation and neutrinoless double beta decay.
  Neutrinos could also play a role in elucidating the nature of dark
  matter and cosmic inflation.

\end{abstract}

\section{Introduction}
\label{sec:introduction}

Neutrinos are the most ubiquitous particles in the universe, over
300/cm$^3$ coming from the Big Bang cross us every second. If
cosmological neutrinos were the only ones available probably there
would be no neutrino physics, given their incredibly tiny interaction
cross sections. Fortunately nature is more generous and stars, such as
our Sun, are copious sources of higher energy neutrinos that can be
detected say, in gigantic underground detectors like Superkamiokande.
Likewise, neutrinos arising from cosmic ray interactions in the upper
atmospheric arrive the Earth from all directions of the sky.
Here too, the agreement between theory and experiments requires the
oscillation hypothesis, characterized by a nearly maximal angle
$\theta^{}_{23}$, surprisingly at odds from expectations based upon
the quark sector.

The resolution of the long-standing discrepancies between theoretical
expectation and experimental measurements of solar and atmospheric
neutrinos has opened this century with a revolution in particle
physics, by providing the first solid evidence for new physics and the
need to revise Standard Model of particle
physics~\cite{Valle:2015pba}.
The latter assembles the fundamental constituents in three generations
of quarks and leptons whose interactions are dictated by the principle
of \SM gauge invariance. It provides a precise theory of particle
interactions, well tested up the highest energies so far explored in
particle accelerators.
While the photon and the gluon, carriers of electromagnetic and the
strong force, are massless, the weak interaction messengers, the W and
the Z are massive, along with all of the quarks and leptons. The basic
theory relies on the principle of gauge invariance and this forbids
mass. The simplest way out is the spontaneous electroweak symmetry
breaking mechanism, which implies the existence of a physical Higgs
boson. Its historic discovery three years ago led many to say that the
\sm is now complete. However, the long-standing discrepancies between
theoretical expectations and experimental measurements of solar and
atmospheric neutrinos requires the existence of neutrino flavor
oscillations~\cite{Forero:2014bxa}, and hence the existence of nonzero
neutrino masses~\cite{Schechter:1980gr}.
This discovery has triggered a revolution in particle physics, as it
provides the first solid evidence for new physics and the need to
revise Standard Model. Indeed, particle physics would have been
``completed'' with the Higgs boson discovery, were it not for the need
to account for neutrino oscillations as well as the cosmological
puzzles such as dark matter, baryon asymmetry and inflation.  In this
talk I will give a brief summary of the current landscape of particle
physics in view of these issues.

\section{Neutrino mixing and oscillations}
\label{sec:oscillations}

The long-standing discrepancy between theoretical expectation and
experimental measurements of solar neutrinos has finally been resolved
in favor of the oscillation mechanism, characterized by an angle
$\theta^{}_{12}$, substantially larger than its CKM analogue, the
Cabbibo angle $\theta^{}_C$.
Similarly, the agreement between theory predictions and measurements
of atmospheric neutrinos at underground experiments, both event yields
and angular distributions, indicates the need for neutrino
oscillations, characterized by a nearly maximal mixing angle
$\theta^{}_{23}$, quite different from its quark sector analogue.

Both solar and atmospheric neutrino discrepancies were crucially
confirmed by the results of Earth-bound experiments based at reactors
and accelerators. For example, the reactor experiment KamLAND pinned
down that oscillations is the mechanism underlying the conversion of
solar neutrinos and identified the relevant region of oscillation
parameters, characterized by a ``small'' angle $\theta^{}_{12}$, as
opposed to oscillations in vacuum. Recent reactor and accelerator
experiments have also provided a good measurement of the third lepton
flavor mixing parameter $\theta^{}_{13}$, with a first hint of
leptonic CP violation just emerging, characterized by a CP phase
$\delta$ which promises to open a new era in neutrino physics.\\


The basic ingredient needed to describe neutrino oscillations is the
lepton mixing matrix $K$, which comes from the mismatch between the
charged and neutral mass matrices arising after the spontaneous
electroweak and lepton number breaking.
If neutrinos get mass \textit{a la seesaw} (see below) then one
expects that the heavy neutrino messengers will couple, subdominantly,
in the charged current weak interaction leading to a rectangular
form~\cite{Schechter:1980gr} for the matrix $K$.

To analyze the current solar, atmospheric, reactor and accelerator
neutrino oscillation data one normally assumes the simplest unitary
form for $K$. The two extra physical CP phases present in $K$ are
called Majorana phases and are most transparently expressed in terms
of the original symmetric parametrization~\cite{Schechter:1980gr}.
However they appear only in neutrinoless double beta decay and other
\lnv processes. Hence they are omitted in neutrino oscillation
analyses, for which the symmetric and the PDG presentations
coincide. 

\begin{figure}[t!]
\centering
\includegraphics[scale=0.4]{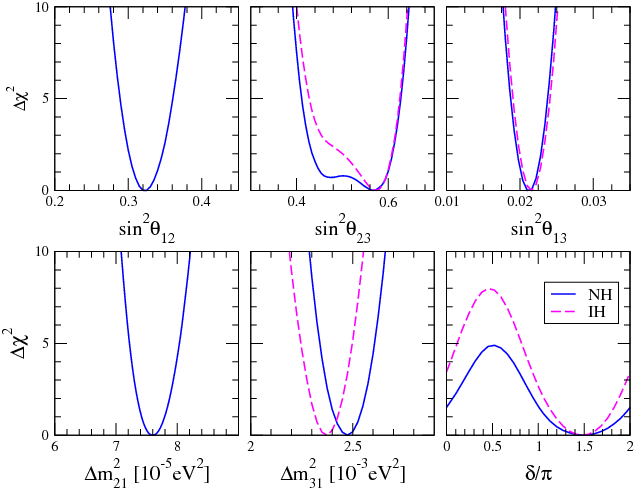}
\caption{Global picture of neutrino oscillation parameters after
  Neutrino-2014, from Ref.~\cite{Forero:2014bxa}.}
\label{fig:osc-fit}
\end{figure}

The summary of the oscillation parameters after the Neutrino 2014
conference are presented in Fig.~\ref{fig:osc-fit} (more discussion in
Lisi's talk).
Clearly one has good determinations of all the oscillation parameters
except for the leptonic CP phase, which is just making its first
appearance in the scene. The squared mass splitting parameters are
tiny, without any counterpart in the charged fermion sector. Likewise,
the values obtained for the solar and atmospheric angles
$\theta^{}_{12}$ and $\theta^{}_{23}$, are much larger than their CKM
counterparts.
The nonzero value of the reactor angle $\theta^{}_{13}$ opens the door
to future leptonic CP violations studies at the upcoming reactor and
accelerator neutrino experiments, such as LBNF-DUNE. The measurement
of the leptonic CP-phase using atmospheric neutrinos has been
discussed in Smirnov's talk.

\section{Effective neutrino mass, seesaw mechanism and unification}
\label{sec:Seesaw}

\begin{figure}[t!]
\centering
\includegraphics[scale=0.4]{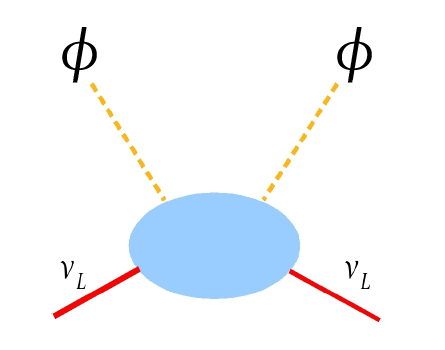}
\caption{Weinberg dimension five \lnv operator leading to neutrino mass.}
\label{fig:Weinberg}
\end{figure}

In the \sm neutrinos are massless so we need new physics in order to
account for neutrino mixing and oscillations. As noted by Weinberg,
one can add non-renormalizable operators, such as the dimension five
operator shown in Fig.~\ref{fig:Weinberg}, that break lepton number
and which would account for the small observed neutrino masses.
We have no clues as to the characteristic scale, the underlying
mechanism or the flavor structure of the relevant operator. If
anything, the neutrino oscillation observations indicate a very
special pattern of mixing parameters, unlikely to be accidental. Its
explanation from first principles, along with the other fermion masses
and mixing parameters, constitutes the so-called \textit{flavor
  problem}, one of the most stubborn problems in particle physics, and
one for which the simplest gauge paradigm falls short at
addressing. Here we stress the challenge of reconciling small CKM
mixing parameters with large lepton mixing angles within a predictive
framework.

The most popular way to induce the operator in Fig.~\ref{fig:Weinberg}
is through the exchange of heavy messengers, as present in SO(10)
Grand unified theories (GUTS). In this case the smallness of neutrino
mass is dynamically explained by minimizing the Higgs potential
through a simple ``1-2-3'' vev (vacuum expectation value) seesaw
relation of the type
 \begin{equation}
   v_3 v_1 \sim {v_2}^2 \:\:\: \text{with~the~hierarchy} \:\:\: v_1 \gg v_2 \gg v_3~. 
 \label{eq:123-vev-seesaw}
 \end{equation}
The isosinglet vev $v_1$ drives the spontaneous breaking of lepton
number symmetry and induces also a small but nonzero isotriplet vev
$v_3$ which generates the $\nu\nu$ entry in the neutrino mass matrix.
Since the isodoublet ${v_2}$ fixes the masses of the weak gauge
bosons, $W$ and $Z$, one sees that $v_3 \to 0$ as $v_1 \to \infty$.
The most popular messengers are heavy ``right-handed'' neutrinos
(type-I seesaw) and heavy triplet scalar with a small induced vev
(type-II seesaw). Although these arise naturally in the framework of
SO(10) GUTS, they may be introduced simply in terms of the \SM
structure.

\section{New physics: to unify or not to unify?}
\label{sec:unify-or-not}

Despite the solid evidence for physics beyond-the-Standard Model in
the neutrino sector, most theoretical extensions, such as grand
unification, have so far been mainly driven by aesthetical principles.
Grand unified theories (GUTS) realize one of the most elegant ideas in
particle physics. The three observed gauge interactions of the
Standard Model which describe electromagnetic, weak, and strong forces
merge into a single one at high energies. GUTS bring a
\textit{rationale} to charge quantization and the quantum numbers of
the Standard Model. They are thought of as an intermediate step
towards the ultimate \textit{theory of everything}, which would also
include gravity. As a generic feature, GUTS break the baryon number
symmetry, allowing protons to decay in many ways. To date, all
attempts to observe proton decay have failed. Here we stress three
attractive features of GUTS:
\begin{itemize}
\item 
Simplest GUTS embed \SM in an enlarged simple Lie group, characterized
by a single unified gauge coupling constant.
\item GUTS open the door to the possibility of relating quark and
  lepton masses.
\item GUTS like SO(10) require the
  existence of right-handed neutrinos and the required breaking of
  B-L implies massive neutrinos.
\end{itemize}

Here we show how non-unified extended electroweak models with massive
neutrinos may unify the gauge couplings as well as relate quark and
lepton masses.

\subsection{Neutrino masses without GUTS}
\label{sec:low-scale-neutrino}

Given that the number and properties of the messengers leading to
Fig.~\ref{fig:Weinberg} are to a large extent arbitrary, one can
devise a variety of low-scale realizations of the seesaw paradigm,
putting it literally ``upside-down''. In particular, the seesaw may be
naturally realized at low-scale, for example, the inverse and the
linear seesaw mechanism. While these can be formulated in a GUT
context~\cite{Malinsky:2005bi}, this is not necessary at
all~\cite{Boucenna:2014zba}.
\begin{figure}[t!]
\centering
\includegraphics[scale=0.9]{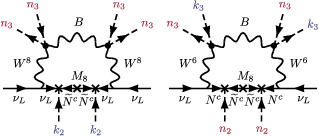}
\caption{Two diagrams which contribute to the light neutrinos mass matrix~\cite{PhysRevD.91.031702}.}
\label{fig:331-octet-rad}
\end{figure}

An alternative low-scale approach to neutrino masses is to assume that
they arise only radiatively, typically as a result of extended
symmetry breaking sectors. However, interesting examples have recently
been suggested where neutrino masses arise from new gauge
interactions, as illustrated in Fig.~\ref{fig:331-octet-rad}. The
crosses denote vev insertions of the relevant scalar multiplets
responsible for symmetry breaking in the relevant extended \TrTrOne
electroweak setup~\cite{PhysRevD.91.031702}.

\subsection{Gauge coupling unification without GUTS}
\label{sec:gauge-coupl-unif}

Within the standard \SM gauge theory gauge coupling unification
constitutes a ``near miss''.  What kind of new physics could make the
gauge coupling constants unify ``exactly''?  The first possibility is
having a full-fledged Grand Unified Theory, as described above. This,
however, entails as phenomenological implication the existence of
proton decay, so far unobserved.

\begin{figure}[h!]
\centering
\includegraphics[scale=0.4]{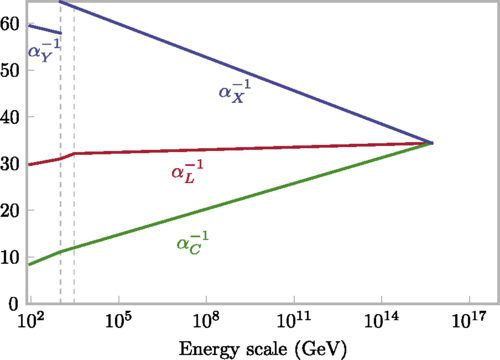}
\caption{Gauge coupling unification in \TrTrOne~model at $3$~TeV, from
  Ref.~\cite{PhysRevD.91.031702}.}
\label{fig:UnificationPlots}
\end{figure}
Alternatively, low energy supersymmetry would provide a simple way to
account for gauge coupling unification.  Such ``completion'' would
however require ``sparticles'' accessible at the LHC, so far not
seen. While we look forward to possible signs of supersymmetry in the
next run of the LHC, we note that the physics responsible for gauge
coupling unification may be \textit{the same inducing small neutrino
  masses}.

A realization of such ``GUT-less'' unification scenario employs the
\TrTrOne electroweak gauge structure, ``explaining'' why there are
three generations from anomaly cancellation~\cite{Singer:1980sw}.
Neutrino masses arise radiatively in the presence of three fermion
octets as illustrated in Fig.~\ref{fig:UnificationPlots}. 
Altogether, one finds that such ``neutrino completion'' scheme unifies
the gauge couplings thanks to the existence of new states providing
neutrino mass. These may lie in the TeV range and hence be accessible
to the LHC.

\subsection{Generalized $b-\tau$  unification without GUTS}
\label{sec:generalized-b-tau}

Flavor symmetries have been suggested as a way to put order in the
``flavor chaos''. Here we stress the striking fact that such
symmetries have the potential of relating quark and charged lepton
masses, in the absence of unification. Indeed, in a class of such \SM
models one can obtain a canonical mass relation~\cite{Bonilla:2014xla}
\begin{equation}\label{eq:MR2}
\frac{m_{b}}{\sqrt{m_{d}m_{s}}}\approx \frac{m_{\tau}}{\sqrt{m_{e}m_{\mu}}}.
\end{equation} 
between down-type quark and charged lepton masses. This formula can be
understood from the group structure, when there are three vacuum
expectation values but only two invariant contractions determining the
Yukawa couplings. Note that Eq.~\ref{eq:MR2} provides a successful
multi-generation generalized b-tau unification scenario which,
moreover, does not require the existence of grand-unification. Note
also that it relates mass ratios instead of absolute masses.

\section{Predicting neutrino oscillation parameters }
\label{sec:pred-neutr-oscill}

A remarkable feature, which came as a surprise, is that the smallest
of the lepton mixing angles is similar to the largest of the CKM
mixing parameters, the Cabibbo angle, while the solar and atmospheric
mixing parameters are rather large~\cite{Forero:2014bxa}.
One phenomenological approach is to take the reactor angle, similar to
the Cabibbo angle, as the universal seed for quark and lepton
mixing. Such \textit{bi-large} schemes point towards Abelian flavor
symmetry groups and Frogatt-Nielsen-type
schemes~\cite{Boucenna:2012xb,Ding:2012wh}.
It has been noted however that the observed neutrino mixing angles
take very special values, atmospheric mixing being nearly {\it
  bi-maximal} with solar mixing nearly {\it tri-maximal}. Hence a {\it
  tri-bimaximal} pattern seems reasonable as a starting
point~\cite{Harrison:2002er}.
Although the full pattern might occur accidentally, it seems that
nature is telling its message here: (i) observations seem to suggest
some symmetry, and (ii) we must also redefine our strategy in flavor
model-building. The challenge is to reconcile the large lepton mixing
with the small CKM parameters in a predictive way.
\begin{figure}[!h]
\centering
\includegraphics[scale=0.5]{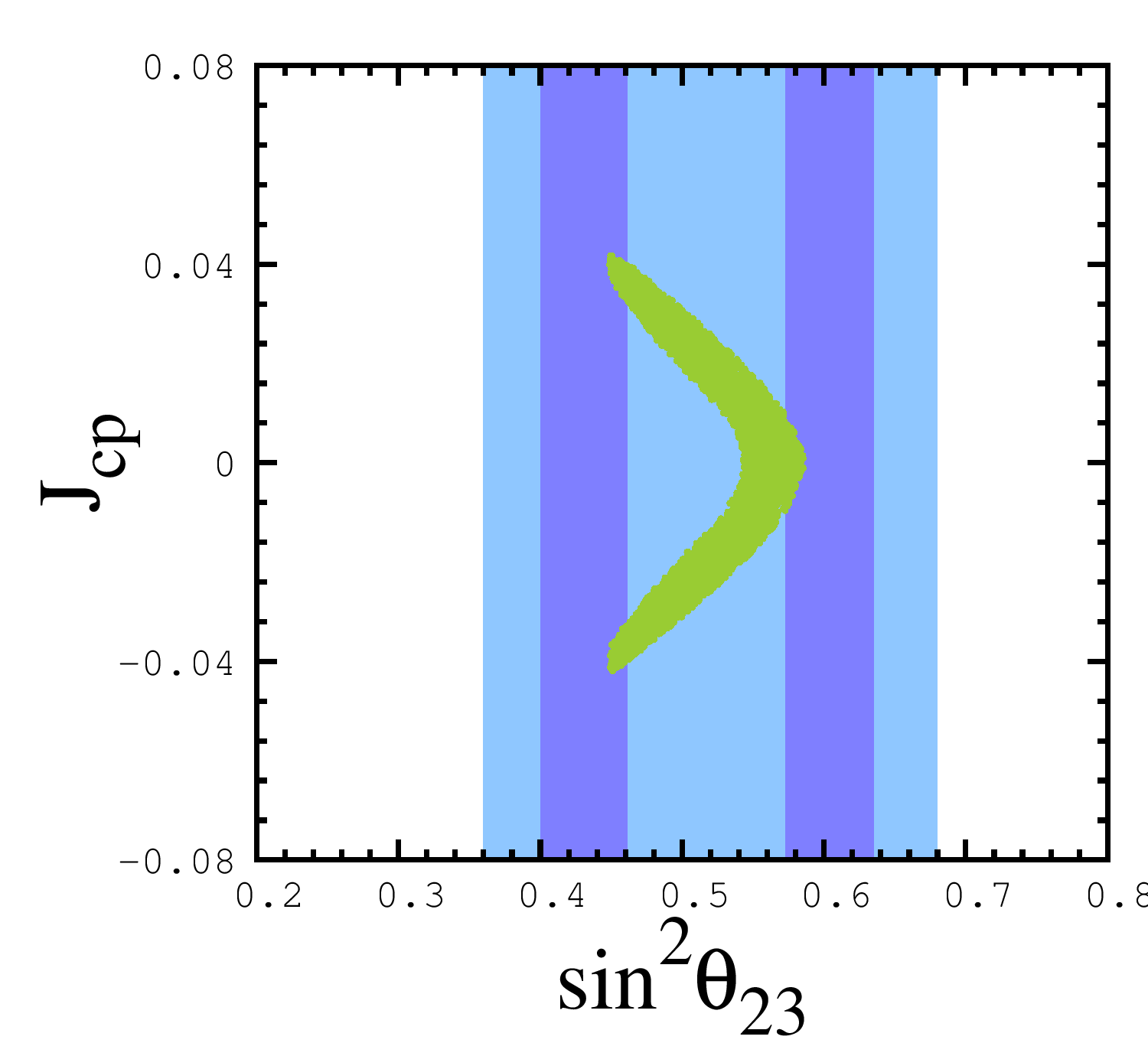}
\caption{Correlating CP violation in neutrino oscillations with the
  octant of the atmospheric mixing parameter $\theta_{23}$, adapted
  from Ref.~\cite{Morisi:2013qna}.}
\label{fig:revamp}
\end{figure}

As a first step one can assign the three lepton families to a
three-dimensional irreducible representation of a non-Abelian flavor
symmetry group, the smallest one being $A_4$.
This opens the way to the possibility of predicting the pattern on
neutrino oscillation parameters.
As simplest zeroth-order predictions one obtains~\cite{Babu:2002dz} a
maximum atmospheric mixing parameter $\theta_{23}=\pi/4$ and zero
reactor angle $\theta_{13}=0$, with a possible prediction as also for
the solar angle, {\it a la tri-bimaximal}.

However recent neutrino oscillation data from reactors and
accelerators measure a nonzero $\theta_{13}$ value, requiring the
early models to be revamped so as to induce a nonzero $\theta_{13}$,
without spoiling the other prediction(s).
This has been done in a minimal way in Ref.~\cite{Morisi:2013qna},
leading to a striking predicted correlation between the magnitude of
CP violation in neutrino oscillations and the octant of the
atmospheric mixing parameter $\theta_{23}$ illustrated in
Fig.~\ref{fig:revamp}.
One sees that, at face value, the left octant necessarily violates
CP. Time will tell whether this predicted \textit{correlations} is
right. Finally we note that flavor-symmetry-based models often predict
the pattern charged \lfv processes~\cite{Morisi:2012fg}.

\section{Neutrinos and electroweak symmetry breaking}
\label{sec:neutr-electr-symm}

After the Higgs boson discovery at CERN it is natural to imagine that
all symmetries in nature are broken spontaneously. It is also
reasonable to imagine that the smallness of neutrino mass is due to
the feeble breaking of lepton number, which can be realized in many
ways, see below. This requires an extension of the \sm Higgs sector
and, if the minimal \SM structure is kept, there must be a physical
Nambu-Goldstone boson, generically called
majoron~\cite{Valle:2015pba}.
\begin{figure}[!h]
  \centering
  \includegraphics[width=0.5\textwidth]{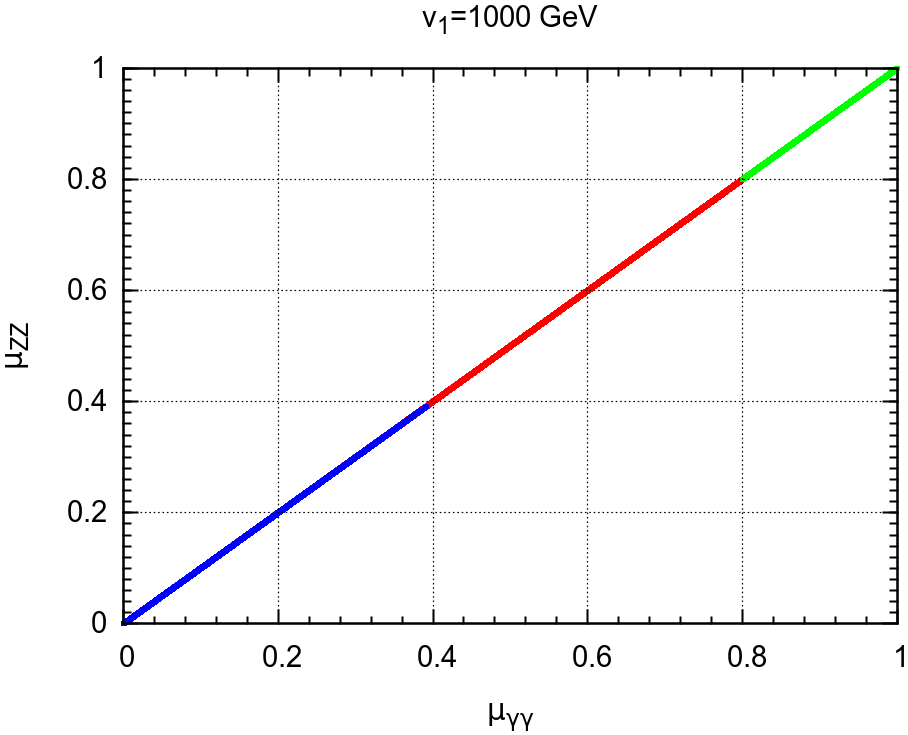}
  \caption{Correlation between $\mu_{ZZ}$ and $\mu_{\gamma\gamma}$. 
  The points in green pass all constraints, from Ref.~\cite{Bonilla:2015uwa}.}
  \label{fig:inv}
\end{figure}

Although the detailed properties of the majoron in general depend on
the model, the existence of new invisible Higgs decays is generically
expected, if \lnv takes place at or below the weak scale. This is easy
to arrange, leading to missing momentum signals at
colliders~\cite{joshipura:1992hp,DeCampos:1996bg,abdallah:2003ry}. Given
the good agreement of the results from ATLAS and CMS with the \sm
Higgs scenario~\cite{Aad:2015zhl} one can place limits on the presence
of invisible channels. Within the simplest \SM spontaneous low-scale
\lnv scenario one finds that the current LHC restrictions on the Higgs
boson decay branchings can be summarized as in Fig.~\ref{fig:inv},
where the parameters $\mu_{ZZ}$ and $\mu_{\gamma\gamma}$ are
``signal-strength'' parameters.
This restriction still leaves an important chunk of Higgs boson mass
and mixing parameters to be explored at the next run of the LHC. Many
alternative richer electroweak breaking sectors leading to the double
breaking of electroweak and lepton number symmetries can be envisaged.

\section{Neutrinoless Double Beta Decay}
\label{sec:neutr-double-beta}

\begin{figure}[t!]
\centering
\includegraphics[scale=0.65]{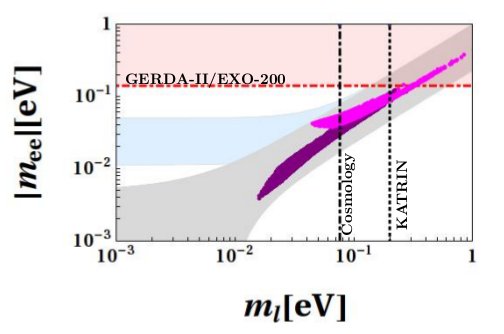}
\caption{Neutrinoless double beta decay effective amplitude parameter
  versus the lightest neutrino mass, in a generic model versus a
  flavor-symmetry-based model, from Ref.~\cite{Bonilla:2014xla}.}
\label{fig:dbd1}
\end{figure}

As we saw neutrino oscillations are insensitive to the absolute
neutrino mass scale.  This can be probed using cosmological data as
well as tritium beta decay endpoint studies~\cite{Valle:2015pba}.  A
specially interesting complementary approach is the search for
neutrinoless double beta decay.  While the two-neutrino double beta
decay has been experimentally observed in many isotopes, so far we
have only experimental lower bounds on the half-lives for the
neutrinoless mode~\cite{Barabash:2011fn}.
However the latter is expected, on general grounds, to take place at
some level, due to the existence of neutrino mass.  Using the previous
oscillation parameters and leaving the values of the Majorana phases
free, one obtains the two broad branches corresponding to the cases of
normal and inverted hierarchies indicated in Fig.~\ref{fig:dbd1}.
The horizontal and vertical lines indicate future expected sensitivities.
Models based upon flavor symmetries often lead, as phenomenological
predictions, to correlations between the neutrino oscillation
parameters. In a large class of such models these translate as lower
bounds for the effective mass parameter $|m_{ee}|$ characterizing
$0\nu\beta\beta$ decay even for the normal mass ordering. This is
seen as the two dark sub-regions in Fig.~\ref{fig:dbd1}. Many other
models leading to a lower bound on the \znbb decay
rate can be constructed~\cite{Dorame:2011eb}.
\begin{figure}[h!]
\centering
\includegraphics[scale=1]{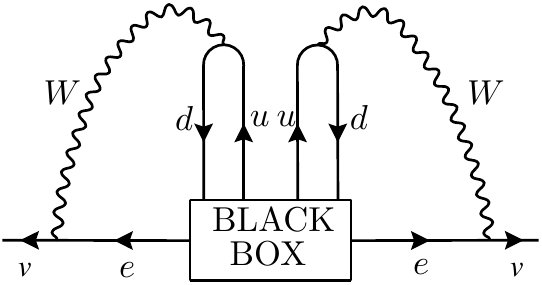}
\caption{Neutrinoless double beta decay implies Majorana neutrinos,
  from Ref.~\cite{Schechter:1981bd}.}
\label{fig:dbd2}
\end{figure}

In gauge theories \znbb can be induced in many ways other than the
neutrino exchange or ``mass mechanism''. For example, there can be
short-range mechanisms involving the exchange of heavy particles such
as present in left-right or supersymmetric extensions of the standard
model~\cite{Bonnet:2012kh,Das:2012ii}.
The significance of neutrinoless double beta decay comes from the fact
that, whatever the mechanism responsible for \znbb in a gauge theory
one can always ``dress'' the corresponding amplitude with W bosons,
showing that a Majorana neutrino mass is necessarily
induced~\cite{Schechter:1981bd}, as illustrated in
Fig.~\ref{fig:dbd2}.  This theorem holds under very general
assumptions, as shown by Lindner and
collaborators~\cite{Duerr:2011zd}.

\section{Neutrinos and cosmology}
\label{sec:neutrinos-affect-cmb}

\begin{figure}[h!]
\centering
\includegraphics[scale=0.15]{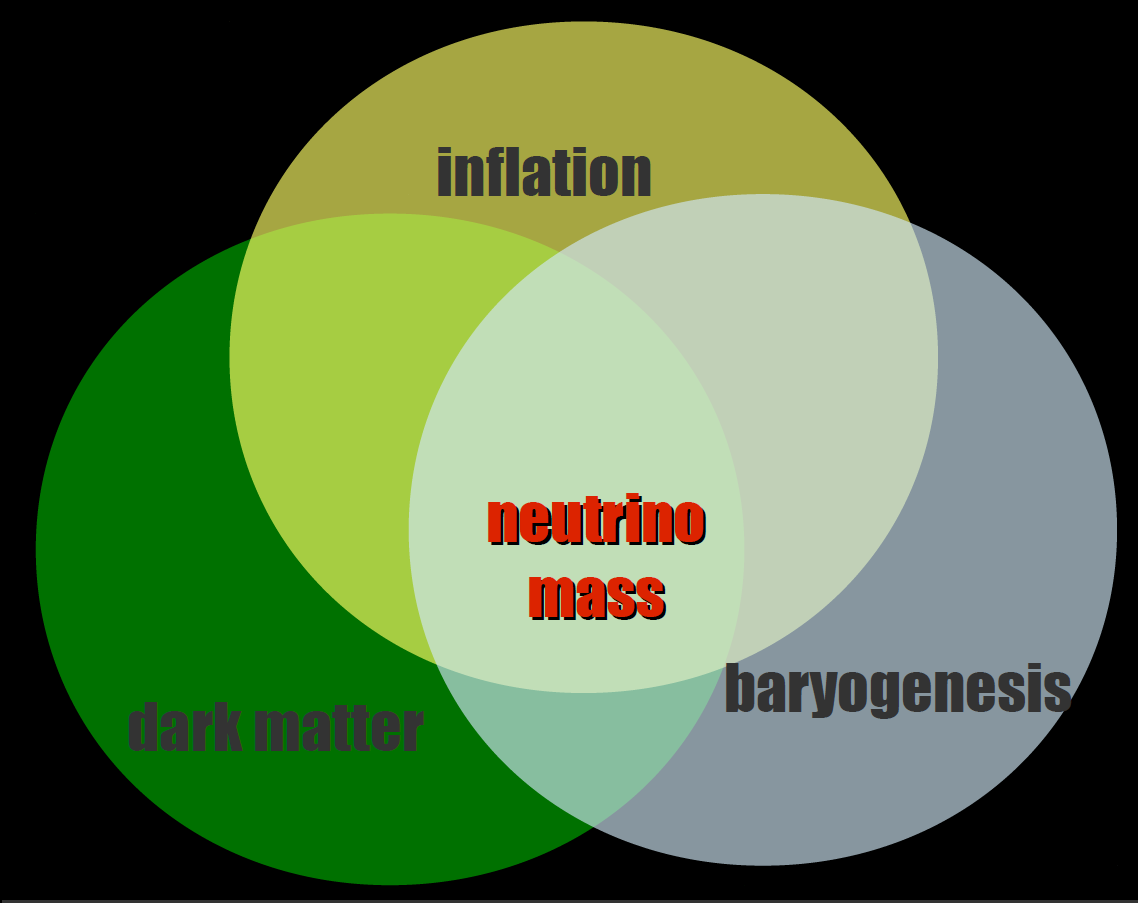}
\caption{Cosmological puzzles possibly associated to the same physics
  that drives neutrino mass.}
\label{fig:joint-cosmo}
\end{figure}

Neutrinos affect the cosmic microwave background (CMB) and large scale
structure in the Universe, playing a key role in the synthesis of
light elements, which takes place when the universe is about a few
minutes old.
The feeble interaction of neutrinos allows us to use them as cosmic
probes, down to epochs far earlier than we can probe with optical
telescopes. The current cosmological puzzles associated with the
baryon number of the universe, inflation and dark matter are probably
associated to epochs earlier than the electroweak phase transition at
$\sim$ 10$^{-12}$sec. It is not inconceivable that (some of) these
puzzles may have a common origin with the physics driving neutrino
masses~\cite{Valle:2015pba}, as schematized in
Fig.~\ref{fig:joint-cosmo}.

Here we focus on the possibility that neutrino masses arise from
spontaneous breaking of ungauged lepton number.
The associated Nambu-Golstone boson may acquire mass from \lnv by
quantum gravity effects at the Planck
scale~\cite{Giddings:1988cx,Banks:2010zn}.
If its mass lies in the keV range, the weakly interacting majoron can
play the role of dark matter particle, providing both the required
relic density as well as the scale for galaxy
formation~\cite{Berezinsky:1993fm}.  Since the majoron couples to
neutrinos proportionally to their tiny mass, it is expected to be very
long-lived, stable on cosmological scales~\cite{Schechter:1981cv}.
Though model-dependent, its coupling to the charged leptons is
expected to be very weak so the majorons produced during the phase
transition may never be in thermal equilibrium during the history of
the universe. Alternatively they could be in thermal equilibrium only
for some period.
\begin{figure}[h!]
\centering
\includegraphics[scale=0.75]{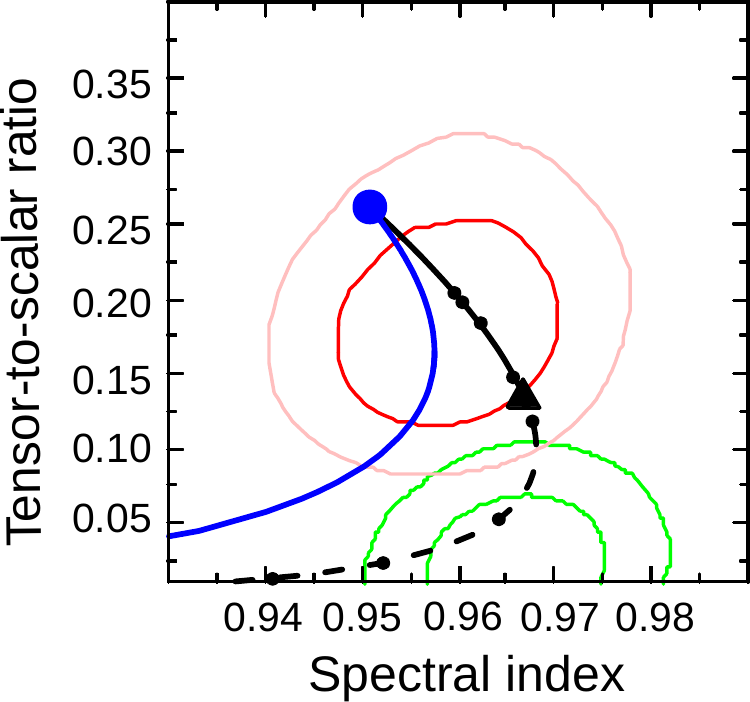}
\caption{Cosmological predictions of seesaw inflation and majoron dark
  matter model of Ref.~\cite{Boucenna:2014uma}.}
\label{fig:biceplanck}
\end{figure}
The lifetime and mass of the late-decaying dark matter majoron
consistent with the cosmic microwave background observations can be
determined~\cite{Lattanzi:2007ux,Lattanzi:2014zla}. Moreover, as a
pseudoscalar, like the $\pi^0$, the dark matter majoron will have a
(sub-dominant) decay to two photons, leading to a characteristic
mono-energetic X-ray emission line~\cite{Lattanzi:2013uza,Lattanzi:2014zla}.
These features fit nicely in models where neutrino masses arise from a
type-II seesaw mechanism.

A recent twist along these lines was the proposal that inflation and
dark matter have a common origin (similar idea was suggested by Smoot
in arXiv:1405.2776 [astro-ph]), with the inflaton identified to the
real part of the complex singlet containing the majoron and breaking
lepton number through its vev~\cite{Boucenna:2014uma}. The resulting
inflationary scenario is consistent with the recent CMB observations,
including the B-mode observation by the BICEP2 experiment re-analized
jointly with the Planck data, as illustrated in
Fig.~\ref{fig:biceplanck}.
The upper (red) contours correspond to the BICEP2 results, while the
lower ones (green) follow from the new analysis released jointly with
PLANCK~\cite{Ade:2015tva}. The lines correspond to 68 and 95\% CL
contours.
Further restrictions on the majoron dark matter scenario should follow from
structure formation considerations.

Work supported by MINECO grants FPA2014-58183-P, Multidark
CSD2009-00064, and the PROMETEOII/2014/084 grant from Generalitat
Valenciana.

\bibliographystyle{utphys}

\providecommand{\href}[2]{#2}\begingroup\raggedright\begin{thebibliography}{}

\end{thebibliography}\endgroup


\begin{thebibliography}{10}

\bibitem{Valle:2015pba} 
  Extensive discussion in 
  J.~W.~F. Valle and J.~C. Romao, {\em {Neutrinos in high energy and
      astroparticle physics}}.  \newblock Wiley-VCH, Berlin, 1st
  edition~ed., 2015.

\bibitem{Forero:2014bxa}
D.~Forero, M.~Tortola, and J.W.F.~Valle, ``{Neutrino oscillations refitted},''
  \href{http://dx.doi.org/10.1103/PhysRevD.90.093006}{{\em Phys.Rev.}
  {\bfseries D90} no.~9, (2014) 093006}
\href{http://arxiv.org/abs/1405.7540}{{\ttfamily arXiv:1405.7540 [hep-ph]}}.

\bibitem{Schechter:1980gr}
J.~Schechter and J.W.F.~Valle, ``{Neutrino Masses in SU(2) x U(1) Theories},''
\href{http://dx.doi.org/10.1103/PhysRevD.22.2227}{{\em Phys.Rev.} {\bfseries
  D22} (1980) 2227}.

\bibitem{Malinsky:2005bi}
M.~Malinsky, J.~Romao, and J.W.F.~Valle, ``{Novel supersymmetric SO(10) seesaw
  mechanism},'' \href{http://dx.doi.org/10.1103/PhysRevLett.95.161801}{{\em
  Phys.Rev.Lett.} {\bfseries 95} (2005) 161801}

\bibitem{Boucenna:2014zba}
S.~M. Boucenna, S.~Morisi, and J.~W.~F. Valle, ``{The low-scale approach to
  neutrino masses},''Adv.\ High Energy Phys.\  {\bf 2014}, 831598 (2014)

\bibitem{PhysRevD.91.031702}
S.~M. Boucenna et. al., 
  ``Small neutrino masses and gauge coupling unification'' {\em Phys. Rev. D}
  {\bfseries 91} (2015) 031702.

\bibitem{Singer:1980sw}
M.~Singer, J.~Valle, and J.~Schechter, ``{Canonical Neutral Current Predictions
  From the Weak Electromagnetic Gauge Group SU(3) X $u$(1)},''
\href{http://dx.doi.org/10.1103/PhysRevD.22.738}{{\em Phys.Rev.} {\bfseries
  D22} (1980) 738}.

\bibitem{Bonilla:2014xla}
  S.~Morisi et. al., 
``Relating quarks and leptons without grand-unification,''
 \textit{ Phys. Rev.} D {\bf 84} (2011) 036003;
 S.~Morisi et. al., 
``Quark-Lepton Mass Relation and CKM mixing in an A4 Extension of the Minimal Supersymmetric Standard Model,''
  \textit{Phys. Rev.} D {\bf 88} (2013) 036001;
  S.~F.~King et, al., 
``Quark-Lepton Mass Relation in a Realistic $A_4$ Extension of the Standard Model,''
 \textit{ Phys. Lett.} B {\bf 724} (2013) 68;
C.~Bonilla   et. al., 
``{Relating quarks and
  leptons with the $T_7$ flavour group},''
  \href{http://dx.doi.org/10.1016/j.physletb.2015.01.017}{{\em Phys.Lett.}
  {\bfseries B742} (2015) 99}

\bibitem{Boucenna:2012xb}
S.~Boucenna, S.~Morisi, M.~Tortola, and J.~Valle, ``{Bi-large neutrino mixing
  and the Cabibbo angle},''
  \href{http://dx.doi.org/10.1103/PhysRevD.86.051301}{{\em Phys.Rev.}
  {\bfseries D86} (2012) 051301}

\bibitem{Ding:2012wh}
G.-J. Ding, S.~Morisi, and J.~Valle, ``{Bilarge neutrino mixing and Abelian
  flavor symmetry},'' \href{http://dx.doi.org/10.1103/PhysRevD.87.053013}{{\em
  Phys.Rev.} {\bfseries D87} no.~5, (2013) 053013}

\bibitem{Harrison:2002er}
P.~Harrison, D.~Perkins, and W.~Scott, ``{Tri-bimaximal mixing and the neutrino
  oscillation data},''
  \href{http://dx.doi.org/10.1016/S0370-2693(02)01336-9}{{\em Phys.Lett.}
  {\bfseries B530} (2002) 167},

\bibitem{Babu:2002dz}
K.~Babu, E.~Ma, and J.~Valle, ``{Underlying A(4) symmetry for the neutrino mass
  matrix and the quark mixing matrix},''
  \href{http://dx.doi.org/10.1016/S0370-2693(02)03153-2}{{\em Phys.Lett.}
  {\bfseries B552} (2003) 207--213},
\href{http://arxiv.org/abs/hep-ph/0206292}{{\ttfamily arXiv:hep-ph/0206292
  [hep-ph]}}.

\bibitem{Morisi:2013qna}
S.~Morisi, D.~Forero, J.~C. Romao, and J.~W.~F. Valle, ``{Neutrino mixing with
  revamped A4 flavour symmetry},'' {\em Phys.Rev.} {\bfseries D88} (2013)
  016003,
\href{http://arxiv.org/abs/1305.6774}{{\ttfamily arXiv:1305.6774 [hep-ph]}}.

\bibitem{Morisi:2012fg}
S.~Morisi and J.~W.~F. Valle, ``{Neutrino masses and mixing: a flavour symmetry
  roadmap},'' {\em Fortsch.Phys.} {\bfseries 61} (2013) 466--492,

\bibitem{Bonilla:2015uwa}
C.~Bonilla, J.~W.~F. Valle, and J.~C. Romão, ``{Neutrino mass and invisible
  Higgs decays at the LHC},''
\href{http://arxiv.org/abs/1502.01649}{{\ttfamily arXiv:1502.01649 [hep-ph]}}.

\bibitem{joshipura:1992hp}
A.~S. Joshipura and J.~W.~F. Valle, ``{Invisible Higgs decays and neutrino
  physics},''
{\em Nucl. Phys.} {\bfseries B397} (1993) 105--122.

\bibitem{DeCampos:1996bg}
F.~de~Campos {\em et~al.}, ``{Searching for Invisibly Decaying Higgs Bosons at
  LEP II},''
{\em Phys. Rev.} {\bfseries D55} (1997) 1316--1325.

\bibitem{abdallah:2003ry}
{\bfseries DELPHI collaboration} Collaboration, J.~Abdallah {\em et~al.},
  ``{Searches for invisibly decaying Higgs bosons with the DELPHI detector at
  LEP},'' {\em Eur. Phys. J.} {\bfseries C32} (2004) 475--492,
\href{http://arxiv.org/abs/hep-ex/0401022}{{\ttfamily hep-ex/0401022}}.

\bibitem{Aad:2015zhl}
{\bfseries ATLAS, CMS} Collaboration, G.~Aad {\em et~al.}, ``{Combined
  Measurement of the Higgs Boson Mass in $pp$ Collisions at $\sqrt{s}=7$ and 8
  TeV with the ATLAS and CMS Experiments},''
\href{http://arxiv.org/abs/1503.07589}{{\ttfamily arXiv:1503.07589 [hep-ex]}}.

\bibitem{Barabash:2011fn}
A.~Barabash, ``{75 years of double beta decay: yesterday, today and
  tomorrow},''
\href{http://arxiv.org/abs/1101.4502}{{\ttfamily arXiv:1101.4502 [nucl-ex]}}.

\bibitem{Dorame:2011eb}
L.~Dorame, D.~Meloni, S.~Morisi, E.~Peinado, and J.~Valle, ``{Constraining
  Neutrinoless Double Beta Decay},''
  \href{http://dx.doi.org/10.1016/j.nuclphysb.2012.04.003}{{\em Nucl.Phys.}
  {\bfseries B861} (2012) 259--270}

\bibitem{Schechter:1981bd}
J.~Schechter and J.~Valle, ``{Neutrinoless Double beta Decay in SU(2) x U(1)
  Theories},''
\href{http://dx.doi.org/10.1103/PhysRevD.25.2951}{{\em Phys.Rev.} {\bfseries
  D25} (1982) 2951}.

\bibitem{Bonnet:2012kh}
F.~Bonnet, M.~Hirsch, T.~Ota, and W.~Winter, ``{Systematic decomposition of the
  neutrinoless double beta decay operator},''
  \href{http://dx.doi.org/10.1007/JHEP03(2013)055,
  10.1007/JHEP04(2014)090}{{\em JHEP} {\bfseries 1303} (2013) 055}

\bibitem{Das:2012ii}
S.~Das, F.~Deppisch, O.~Kittel, and J.~Valle, ``{Heavy Neutrinos and Lepton
  Flavour Violation in Left-Right Symmetric Models at the LHC},''
  \href{http://dx.doi.org/10.1103/PhysRevD.86.055006}{{\em Phys.Rev.}
  {\bfseries D86} (2012) 055006},
\href{http://arxiv.org/abs/1206.0256}{{\ttfamily arXiv:1206.0256 [hep-ph]}}.

\bibitem{Duerr:2011zd}
M.~Duerr, M.~Lindner, and A.~Merle, ``{On the Quantitative Impact of the
  Schechter-Valle Theorem},'' {\em JHEP} {\bfseries 1106} (2011) 091

\bibitem{Giddings:1988cx}
S.~B. Giddings and A.~Strominger, ``{Loss of Incoherence and Determination of
  Coupling Constants in Quantum Gravity},''
\href{http://dx.doi.org/10.1016/0550-3213(88)90109-5}{{\em Nucl.Phys.}
  {\bfseries B307} (1988) 854}.

\bibitem{Banks:2010zn}
T.~Banks and N.~Seiberg, ``{Symmetries and Strings in Field Theory and
  Gravity},'' \href{http://dx.doi.org/10.1103/PhysRevD.83.084019}{{\em
  Phys.Rev.} {\bfseries D83} (2011) 084019},
\href{http://arxiv.org/abs/1011.5120}{{\ttfamily arXiv:1011.5120 [hep-th]}}.

\bibitem{Berezinsky:1993fm}
V.~Berezinsky and J.~Valle, ``{The KeV majoron as a dark matter particle},''
  \href{http://dx.doi.org/10.1016/0370-2693(93)90140-D}{{\em Phys.Lett.}
  {\bfseries B318} (1993) 360--366},
\href{http://arxiv.org/abs/hep-ph/9309214}{{\ttfamily arXiv:hep-ph/9309214
  [hep-ph]}}.

\bibitem{Schechter:1981cv}
J.~Schechter and J.~Valle, ``{Neutrino Decay and Spontaneous Violation of
  Lepton Number},''
\href{http://dx.doi.org/10.1103/PhysRevD.25.774}{{\em Phys.Rev.} {\bfseries
  D25} (1982) 774}.

\bibitem{Boucenna:2014uma}
S.~Boucenna, S.~Morisi, Q.~Shafi, and J.~Valle, ``{Inflation and
  majoron dark matter in the seesaw mechanism},''
  \href{http://dx.doi.org/10.1103/PhysRevD.90.055023}{{\em Phys.Rev.}
  {\bfseries D90} (2014) 055023}

\bibitem{Lattanzi:2007ux}
M.~Lattanzi and J.~Valle, ``{Decaying warm dark matter and neutrino masses},''
  \href{http://dx.doi.org/10.1103/PhysRevLett.99.121301}{{\em Phys.Rev.Lett.}
  {\bfseries 99} (2007) 121301},
\href{http://arxiv.org/abs/0705.2406}{{\ttfamily arXiv:0705.2406 [astro-ph]}}.

\bibitem{Lattanzi:2014zla}
M.~Lattanzi, S.~Riemer-Sørensen, M.~Tortola, and J.~Valle, ``{Constraints on
  majoron dark matter from cosmic microwave background and astrophysical
  observations},''
\href{http://dx.doi.org/10.1016/j.nima.2013.09.009}{{\em Nucl.Instrum.Meth.}
  {\bfseries A742} (2014) 154--157}.

\bibitem{Lattanzi:2013uza}
M.~Lattanzi, S.~Riemer-Sorensen, M.~Tortola, and J.~W.~F. Valle, ``{Updated CMB
  and x- and $\gamma$-ray constraints on Majoron dark matter},''
  \href{http://dx.doi.org/10.1103/PhysRevD.88.063528}{{\em Phys.Rev.}
  {\bfseries D88} (2013) 063528},
\href{http://arxiv.org/abs/1303.4685}{{\ttfamily arXiv:1303.4685
  [astro-ph.HE]}}.

\bibitem{Ade:2015tva}
{\bfseries BICEP2, Planck} Collaboration, P.~Ade {\em et~al.}, ``{Joint
  Analysis of BICEP2/$Keckarray$ and $Planck$ Data},''
  \href{http://dx.doi.org/10.1103/PhysRevLett.114.101301}{{\em Phys.Rev.Lett.}
  {\bfseries 114} (2015) 101301}

\end{thebibliography}

\providecommand{\href}[2]{#2}\begingroup\raggedright\endgroup
\end{document}